\documentclass[prx,twocolumn,english,superscriptaddress,floatfix,longbibliography]{revtex4-2}
\usepackage{amsmath}
\usepackage{amssymb}
\usepackage{float}
\usepackage{caption}
\usepackage{subcaption}
\usepackage{svg}
\usepackage{quantikz}
\usepackage{graphicx}
\usepackage{dcolumn}
\usepackage{bm}
\usepackage{hyperref}

\usepackage{ragged2e}
\makeatletter
\renewcommand{\@makecaption}[2]{%
  \setbox\@tempboxa\hbox{\small \textbf{#1.} #2}%
  \ifdim \wd\@tempboxa >\hsize
    \small \textbf{#1.} \justifying #2\par
  \else
    \hbox to\hsize{\small \textbf{#1.} \justifying #2\hfil}%
  \fi}
\makeatother

\begin{document}

\preprint{APS/123-QED}

\title{Protocol for Efficient Generation of Fusion-Based Quantum Computing Resource States from Quantum Emitters}

\author{Nishad Manohar}
\email{nishadm2@illinois.edu}
\author{Arshag Danageozian}
\author{Evangelia Takou}
\altaffiliation[Current address: ]{Duke Quantum Center, Duke University, Durham, North Carolina 27701, USA and Department of Electrical and Computer Engineering, Duke University, Durham, North Carolina 27708, USA}
\author{Edwin Barnes}
\author{Sophia E. Economou}
\email{economou@vt.edu}
\affiliation{%
Department of Physics, Virginia Tech, Blacksburg, Virginia 24061, USA
}%
\affiliation{Center for Quantum Information Science and Engineering, Virginia Tech, Blacksburg, Virginia, 24061, USA}

\date{\today}

\begin{abstract}
Fusion-based quantum computing (FBQC) relies on a set of small, typically photonic, resource states that are fused together through Bell state measurements. The main bottleneck of FBQC is the low rate of generating the resource states, which stems from the probabilistic nature of photonic fusion gates. Previous work introduced a general algorithm for constructing circuits that deterministically generate photonic resource states from a minimal number of quantum emitters for a specified photon emission ordering. However, finding the minimal number of emitters and CNOT gates across all possible orderings is an NP-hard problem. Here, we exploit the symmetries present in FBQC resource states to dramatically simplify this optimization problem. We find that logically encoded 24-photon FBQC resource states can be produced using as few as 3 emitters and 11 CNOTs.
\end{abstract}

\maketitle


\section{\label{sec:level1}Introduction}
Measurement-based quantum computation (MBQC) is perhaps the most promising paradigm for photonic quantum computing \cite{raussendorf2001one, raussendorf2003measurement}. The primary challenge in MBQC is the need to create large, entangled photonic resource states efficiently and with high fidelity. This challenge is much less daunting in a variant of MBQC called fusion-based quantum computation (FBQC) ~\cite{FBQC}, where the size of the resource states is constant and independent of the size of the computation, unlike in the original MBQC. Once generated, these resource states are then fused \cite{browne2005resource} in a fusion network to perform the desired computation. Moreover, implementations of FBQC require fewer reconfigurable components, which makes it appealing for integrated photonic platforms. Consequently, FBQC has become a promising approach to achieving fault-tolerant quantum computation.

Since photons do not interact directly, the required resource states for photonic quantum computation are traditionally generated via single-photon sources and linear optics~\cite{resourcestategeneration, kok2007linear}. However, due to the probabilistic nature of fusion gates between photonic qubits \cite{browne2005resource} (stemming from the $50\%$ success probability of a photonic Bell state measurement), the generation of photonic FBQC resource states is inherently probabilistic, with a success probability that decays exponentially with the number of photons in the state. This is especially problematic when more photons are used to build logically encoded states that protect against photon loss, e.g., by implementing graph codes \cite{bell2023optimizing, pettersson2025long} or the Shor code \cite{pankovich2024high, song2024encoded, wein2025minimizing}, as proposed in the original FBQC paper \cite{FBQC}. So, although FBQC requires only constant-size resource states, generating these states at a sufficiently high rate remains a fundamental bottleneck for this paradigm of quantum computing.

A promising route to overcoming this challenge is to replace probabilistic linear-optical methods with deterministic resource-state generation schemes based on quantum emitters. The general principles behind this approach were first put forward in Ref. ~\cite{schon2005sequential}. This was followed by a specific protocol for generating one-dimensional photonic cluster states from a single, spinful quantum emitter in the form of an electron confined to an optically active quantum dot ~\cite{LinderRudolph}. This was later demonstrated experimentally in quantum dot platforms \cite{schwartz2016deterministic, istrati2020sequential, cogan2023deterministic, coste2023high, huet2025deterministic}. Shortly after the scheme in Ref.~\cite{LinderRudolph} was proposed, it was extended to 2D cluster states using two coupled emitters \cite{economou2010optically,gimeno-segovia2019}. Neutral atom platforms have also had great success in the generation of photonic cluster states \cite{thomas2022efficient, thomas2024fusion}, including the generation of the largest 1D optical photonic cluster state to date, comprised of 12 photons with demonstrated entanglement \cite{thomas2022efficient}. The implementation of fusion alongside deterministic generation protocols has also been demonstrated in both platforms \cite{thomas2024fusion, meng2025temporal}. This hybrid approach is more efficient than fully linear-optical and fusion methods, since it relies on the fusion of deterministically generated ``building-block'' states, known as caterpillar states \cite{hilaire2023near}, which are encoded versions of 1D cluster states that can be fused together to form FBQC resource states \cite{wein2025minimizing}.

Recently, an algorithm was devised that takes an arbitrary graph state as input and outputs generation circuits that utilize a minimal number of quantum emitters \cite{BikunsAlgorithm}. This algorithm requires specifying an emission ordering of the photons in the target graph state, as the minimal number of emitters needed to produce the state depends on this ordering. In applications where the emission ordering is fixed by other considerations (e.g., if the photons are to be measured in the order they are produced to minimize photon storage requirements), then the heuristic algorithms of Ref.~\cite{takou2025optimization} can be used to reduce the emitter-emitter CNOT count. However, in situations where the emission ordering is unimportant, or at least less important than minimizing the number of emitters and CNOTs, one would then want to find the optimal orderings that minimize these resources. However, this optimization problem maps to a known NP-hard graph theory problem called the linear rank-width problem~\cite{BikunsAlgorithm}. Even though the FBQC resource states are rather small, the number of orderings grows factorially with the size of the state; an unencoded 6-photon ring state has 6! = 720 orderings, while the 24-photon encoded ring state of Ref.~\cite{FBQC} has 24! = $6\times 10^{23}$ orderings. The latter is far too large for a brute-force search for optimal orderings to be feasible.

In this article, we exploit the symmetries of FBQC resource states to efficiently find near-optimal generating circuits that utilize a small number of quantum emitters. We show how the symmetry of the resource state reduces the size of the search space, allowing us to search over emission orderings efficiently. While a direct implementation of this symmetry-based reduction strategy still leads to an intractably large search space in the case of the 24-photon encoded state, we show that a slightly modified strategy circumvents this issue, allowing us to obtain near-optimal generation circuits in this case as well. These circuits require only 3 emitters and 11 emitter-emitter CNOT gates to implement. Although we focus here on $n$-ring graph states and their logically encoded counterparts, our approach can be generalized to the generation of other photonic resource states, e.g., for quantum sensing or communications.

The article is organized as follows: In Section~\ref{sec:background}, we briefly review the algorithm proposed in Ref.~\cite{BikunsAlgorithm}, which we utilize in our work. Section~\ref{sec: results} describes the methodology that we follow and showcases our results for both the unencoded (Sec.~\ref{sec:results_unencoded}) and encoded (Sec.~\ref{sec:results_encoded}) resource states. Section~\ref{sec:bounds} then complements our approach by presenting theoretical upper bounds on the total number of required emitter-emitter CNOTs. Finally, Section~\ref{sec:conclusion} provides a discussion and summary of our results.

\section{Background} \label{sec:background}
The algorithm we utilize in this work for the generation of photonic graph states from quantum emitters is the circuit-solving algorithm developed in Ref.~\cite{BikunsAlgorithm}. It relies on the stabilizer formalism and the tableau representation of stabilizer states \cite{Stabilizer,Tableau} to classically simulate the entire state generation process. Here we briefly discuss this technique.

Given a target photonic graph state $\vert G \rangle$ with $n_{p}$ photons, and a minimal number $n_e$ of quantum emitters (also determined by the algorithm of Ref.~\cite{BikunsAlgorithm} for a fixed emission ordering), the algorithm works by solving for the generating circuit in reverse, starting from the target photonic graph state and decoupled emitters, i.e., $\ket{G}\ket{0}^{\otimes n_e}$, and working backward to the state $\ket{0}^{\otimes n_p}\ket{0}^{\otimes n_e}$, which represents decoupled emitters and no photons.

The first step of the algorithm is to find the minimum number $n_{e}$ of required emitters in the generating circuit. This is accomplished by computing the so-called height function (which characterizes the bipartite entanglement entropy in the target graph state as a function of the bi-partition point when the graph is arranged in a 1D lattice in accordance with the chosen photon emission ordering) using the row-reduced echelon form (RREF) gauge choice for the stabilizers~\cite{RREF}. As shown in Ref.~\cite{BikunsAlgorithm}, the maximum value of the height function corresponds to the minimum number of required emitters.

The next step is to solve for the circuit itself by evolving backwards in time step by step from the target state $\ket{G}\ket{0}^{\otimes n_e}$ to the initial state $\ket{0}^{\otimes n_p}\ket{0}^{\otimes n_e}$. The algorithm iteratively builds the time-reversed circuit, utilizing the height function of the state at each iteration to decide the next set of operations to include. Possibilities include photon emission, emitter measurements, and Clifford gates. Every emission is treated as an emitter-photon CNOT (where the photon is initially in the $\ket{0}$ state prior to emission) and thus is described by the stabilizer formalism. The algorithm assumes the ability to perform only single-qubit Clifford gates on the photons, and the only interaction the photon has with the emitters is the emission process itself. In addition, the algorithm allows any Clifford gate to be performed on and between emitters. For further details, we refer the reader to the original paper \cite{BikunsAlgorithm}.

Our goal is to use this algorithm to efficiently search for circuits that generate the resource states for FBQC, with the goal of finding circuits that require the fewest emitters and CNOTs. Since the definition of ``resource state" for FBQC is very broad, we will focus on the example states laid out in Ref.~\cite{FBQC}, particularly on the hexagonal graphs shown in Fig.~\ref{fig:6qubitring}, as well as their logically encoded generalizations. We note that most of the methods we develop for these graphs can be generalized to other graph states with a limited number of photons and discrete symmetries.

In what follows, when counting the number of CNOTs, we exclusively refer to the CNOTs \textit{between emitters}. Namely, we do not include the emission of the photons as part of the CNOT count, as this is just a constant number ($n_{p}$), independent of the emission ordering, and photon emission is far less costly than an actual two-emitter CNOT gate in experiments.

\section{Methods and Results} \label{sec: results}
As mentioned earlier, finding the globally optimal emission orderings and associated generating circuits for a target graph state is NP-hard. This would require searching across all factorially many possibilities. However, we will show that by utilizing the symmetries of the target FBQC resource states, we can achieve a reduction in the computational cost of the search. The features we use are (\textit{i}) symmetry under certain graph transformations when the vertices (emitted photons) are ordered by emission time, which we use to reduce the
search space by only searching over non-equivalent emission orderings, and (\textit{ii}) graph truncation, which allows
us to perform a brute-force search more efficiently after the symmetry-enabled reduction. In what follows, we explain our approach and its implications for the generation of both the unencoded and encoded versions of the FBQC resource states.

\subsection{Unencoded $n$-Ring Graph States} \label{sec:results_unencoded}
\begin{figure}[!htbp]
\centering
     \begin{subfigure}[b]{0.225\textwidth}
         \centering
         \includegraphics[width=\textwidth]{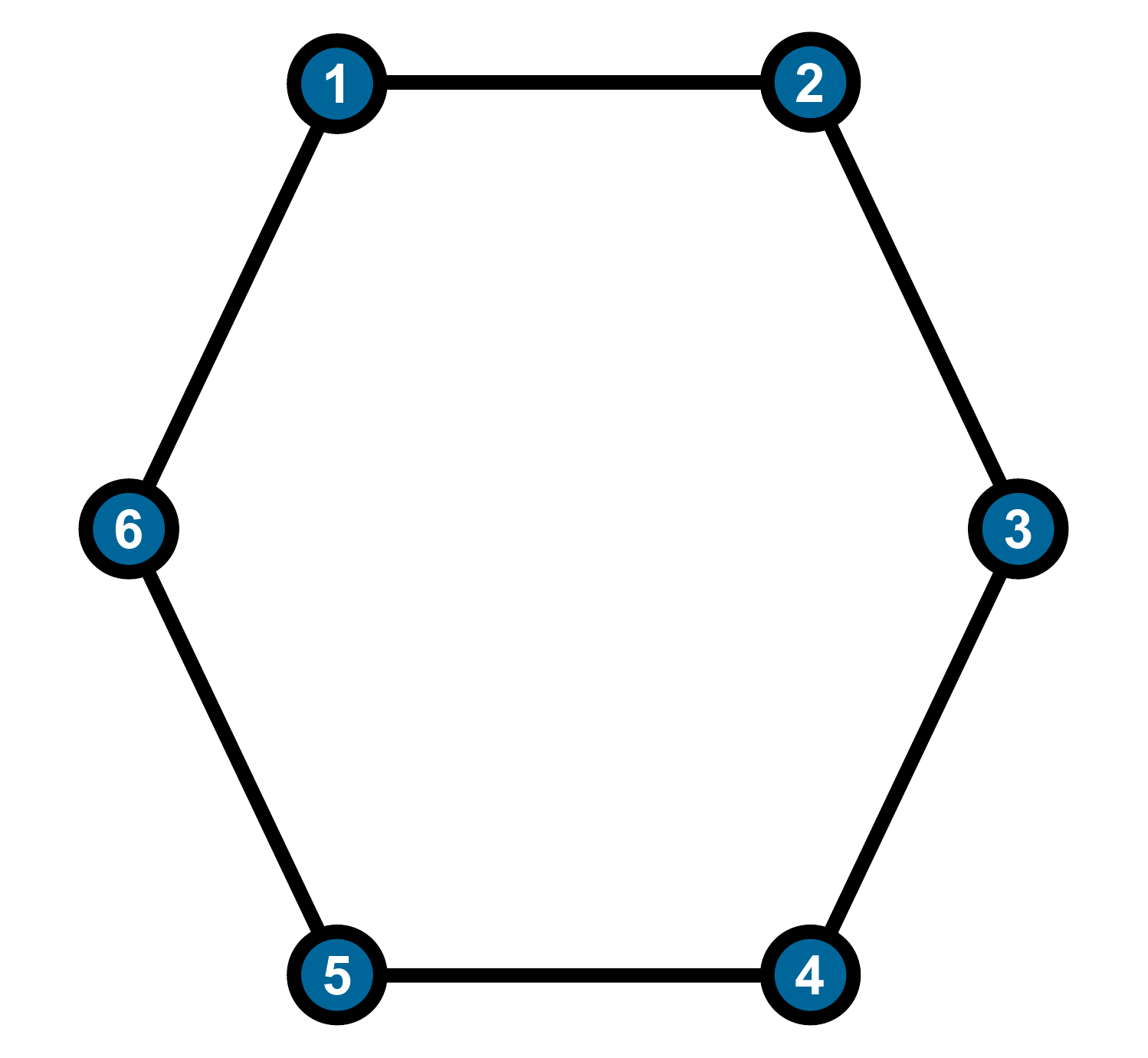}
         \caption{6-qubit ring with original emission ordering}
         \label{fig:6qubitring}
     \end{subfigure}
     \vfill
     \centering
     \begin{subfigure}[b]{0.225\textwidth}
         \centering
         \includegraphics[width=\textwidth]{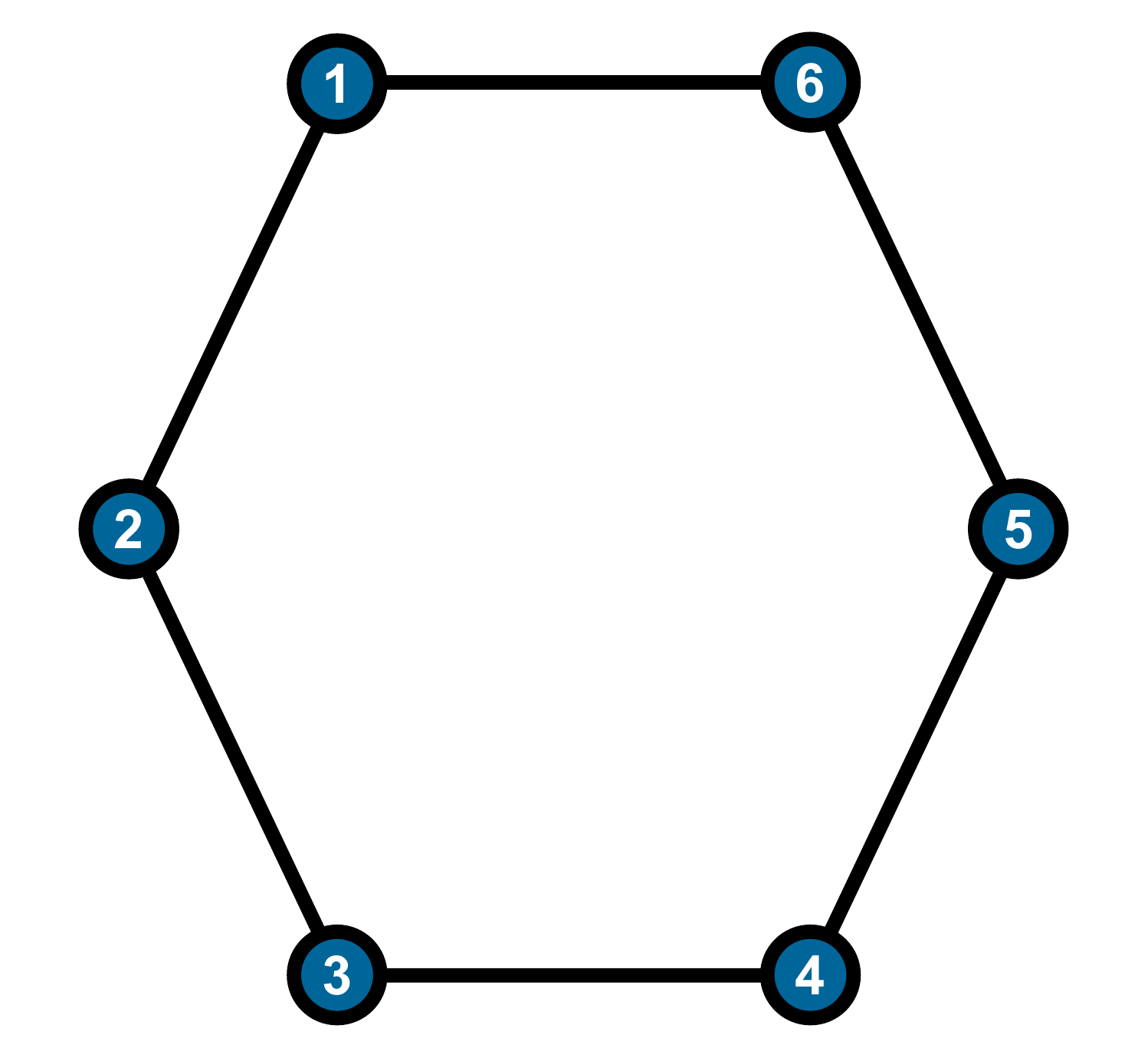}
         \caption{6-qubit ring with reflected emission ordering}
         \label{fig:6ref}
     \end{subfigure}
     \hfill
     \begin{subfigure}[b]{0.225\textwidth}
         \centering
         \includegraphics[width=\textwidth]{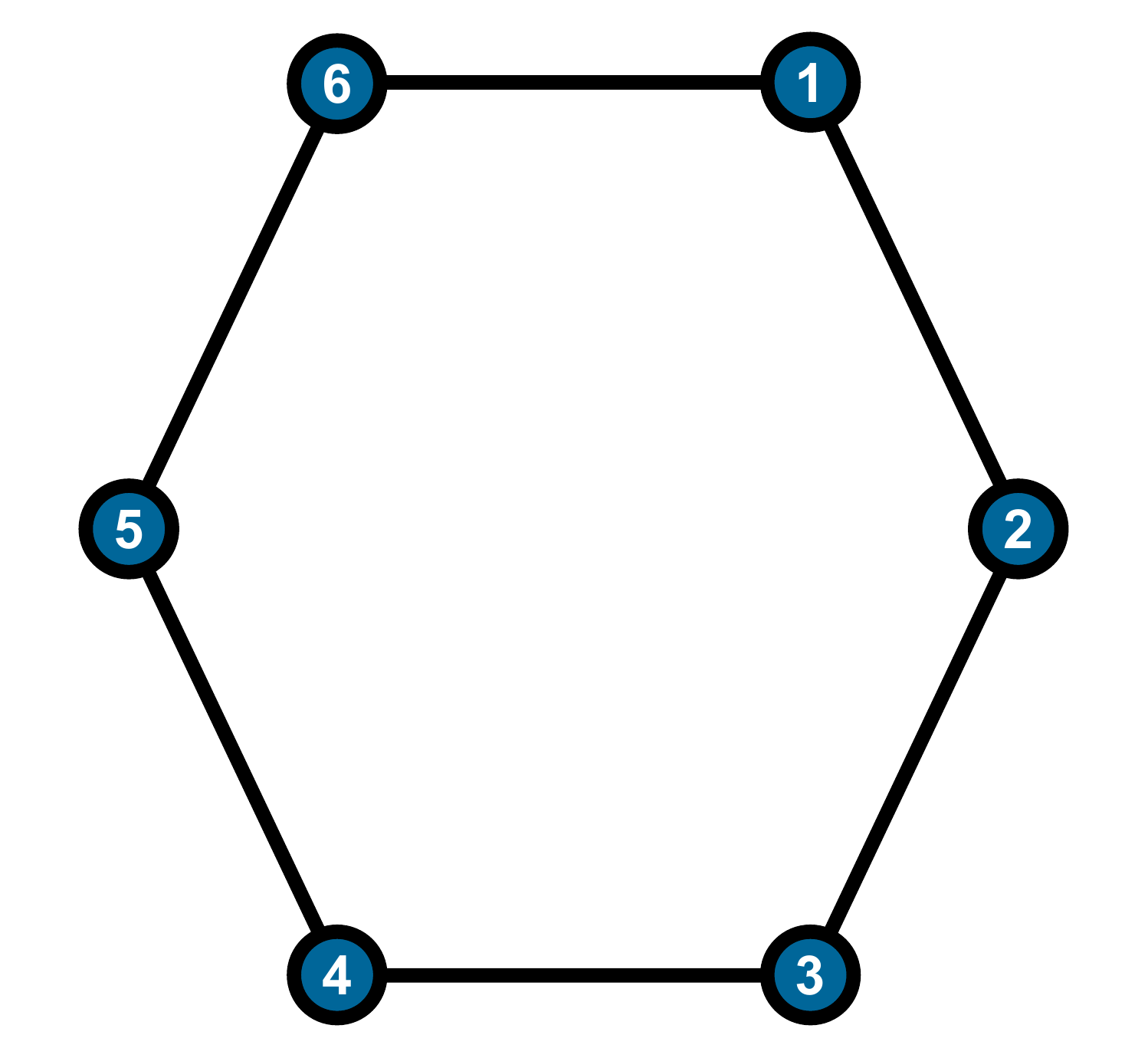}
         \caption{6-qubit ring with rotated emission ordering}
         \label{fig:6rot}
     \end{subfigure}
        \caption{\textit{Resource states with identical generating circuits}: The $D_{12}$ symmetry of the 6-ring graph implies that the target state with emission orderings specified in (a), (b), and (c) must have identical generating circuits.}
        \label{fig:three graphs}
\end{figure}

To start, we quantify the size of the search space, both to understand the problem and to justify some of the heuristics and strategies discussed later in this section. Our starting point is to consider the simplest resource states for FBQC, ring graph states. Assume that we have an $n$-qubit ring graph with a labeled emission ordering, as Fig.~\ref{fig:6qubitring} illustrates for the 6-ring resource state. Naively, there are $n!$ emission orderings; however, the symmetry of the graph state reduces the size of this search space. To better demonstrate this, consider the two labeled rings in Figs.~\ref{fig:6ref}, ~\ref{fig:6rot}. The emission orderings presented are different permutations of Fig.~\ref{fig:6qubitring}. However, due to the rotational and reflective symmetry of the graph, they yield the same graph state as Fig.~\ref{fig:6qubitring}. From this observation, we can also show that the number of ``equivalent orderings" for an $n$-qubit ring state is given by the cardinality of the dihedral group $|D_{2n}| = 2n$. 


Therefore, the size of the search space is given by 
\begin{equation}
N = \dfrac{n!}{|D_{2n}|} = \dfrac{n!}{2n} = \dfrac{(n-1)!}{2} \; .
\label{eq:one}
\end{equation}
From this, the $6$-qubit ring yields a search space of 60 emission orderings, which is significantly less than the naive counting of $6!=720$.

\begin{figure}[!htbp]
    \centering
    \includegraphics[width=0.5\textwidth]{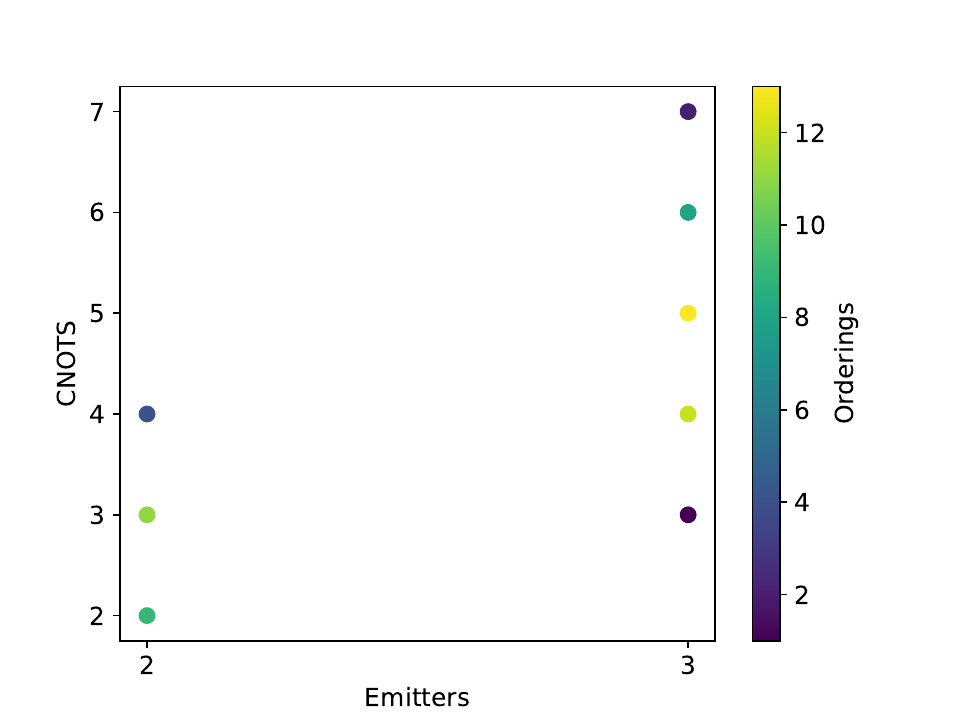}
    \caption{Emitter vs CNOT count for the 6-qubit ring state. The color bar indicates the number of distinct emission orderings for each (Emitters, CNOTs) pair.}
    \label{fig:6QubitScatter}
\end{figure}

\begin{figure}[!htbp]
    \centering
    \includegraphics[width=0.5\textwidth]{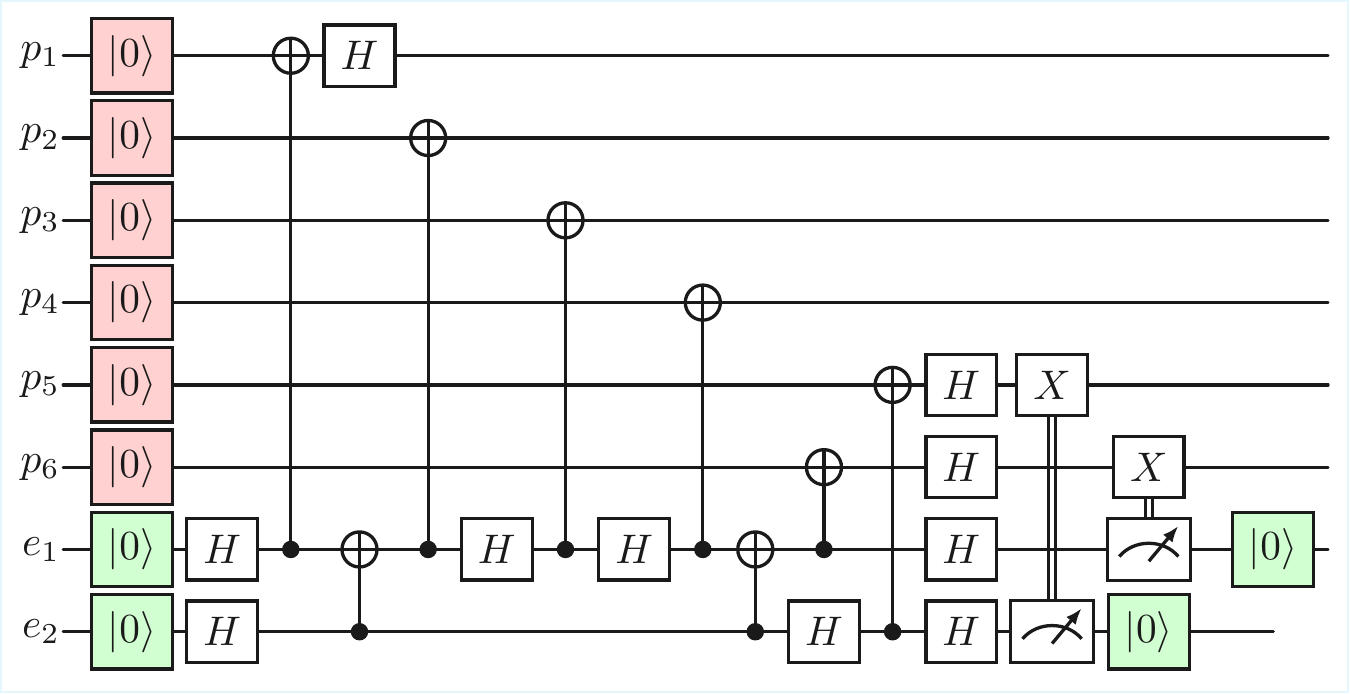}
    \caption{Optimized generating circuit for the 6-qubit ring graph state for edgelist [[1, 2], [1, 4], [2, 3], [3, 6], [4, 5], [5, 6]], corresponding to the emission ordering [1, 2, 3, 6, 5, 4] going around the 6-ring.}
    \label{fig:6QubitCircuit}
\end{figure}

We perform a brute-force search in this reduced space to find the optimal generating circuits for all non-equivalent emission orderings. For each unique emission ordering, we find the minimum number of emitters and the emitter-emitter CNOT count, following the algorithm proposed in Ref.~\cite{BikunsAlgorithm}. Fig.~\ref{fig:6QubitScatter} shows a distribution of emitter vs CNOT counts for different emission orderings.
As we can see, the optimal circuit for generating a 6-qubit ring requires only two emitters and two CNOTs between emitters. We find 9 distinct circuits with an optimal number of CNOTs (equal to 2); one such circuit is given in Fig.~\ref{fig:6QubitCircuit}. The corresponding distinct emission orderings around the 6-ring are given as follows: [1, 3, 2, 4, 5, 6],
[1, 2, 4, 5, 6, 3],
[1, 2, 3, 6, 5, 4],
[1, 2, 3, 4, 6, 5],
[1, 3, 2, 6, 4, 5],
[1, 2, 5, 4, 6, 3],
[1, 4, 2, 6, 3, 5],
[1, 4, 3, 6, 2, 5], and
[1, 5, 2, 4, 3, 6].

However, because photon loss is the dominant source of noise in FBQC, it is not sufficient to generate the unencoded 6-ring graph states. In fact, to achieve fault tolerance, these resource states must be appropriately encoded to counter photon loss and failure of fusion measurements \cite{FBQC, bell2023optimizing, pankovich2024high, song2024encoded}. We discuss the efficient generation of logically encoded states next.

\subsection{Encoded $n$-Ring Graph States} \label{sec:results_encoded}

\begin{figure}[!htbp]
    \centering
    \includegraphics[width=0.3\textwidth]{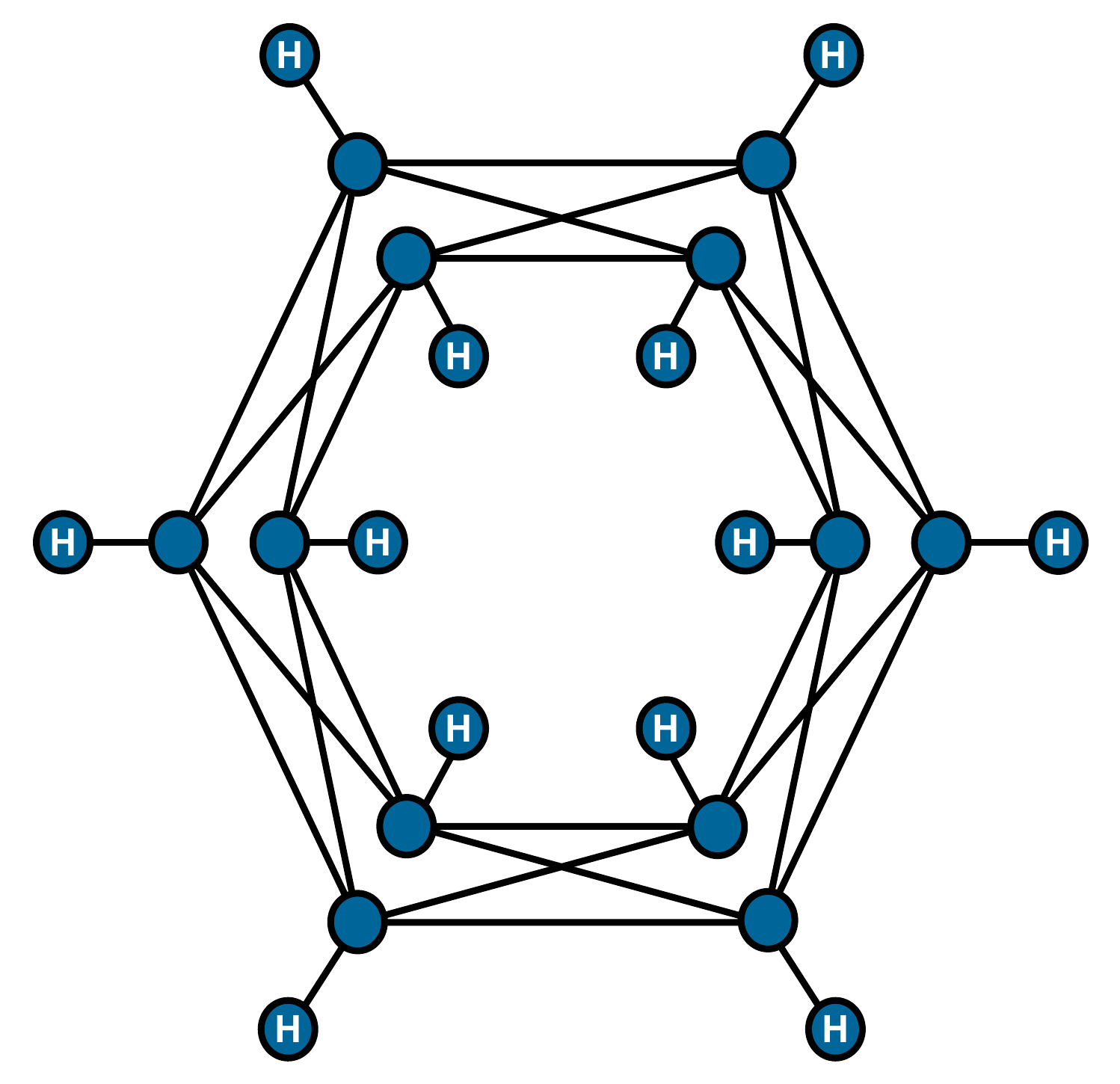}
    \caption{\textit{Encoded Resource State for FBQC}: 24-qubit, (2,2) Shor-encoded, 6-ring graph state. Here, $H$ denotes Hadamard gates on the corresponding leaf photons.}
    \label{fig:24qubitgraph}
\end{figure}

As discussed in the original proposal for FBQC \cite{FBQC}, a better resource is the (2,2) Shor-encoded version of the 6-qubit ring state, as shown in Fig.~\ref{fig:24qubitgraph}. This code is also known as the [[4, 1, 2]] CSS code \cite{gottesman1997stabilizer}, which requires using four physical photons to encode each logical photonic qubit in the original 6-ring graph state. More generally, the (2,2) Shor-encoded version of an $n$-qubit ring will have $4n$ physical qubits.

The increase in the number of physical qubits required for an encoded ring state (and hence the number of emission orderings) leads to an exponentially larger search space. As before, after accounting for the discrete symmetries of the target graph, the total size of the search space for $n>4$ is given by
\begin{eqnarray}
    N = \dfrac{(4n)!}{2n\cdot 2^{n}} \; ,
    \label{eq:totalshorsize}
\end{eqnarray}
which is computed by the orbit-stabilizer theorem~\cite{dummitfoote}. To see how, consider the group of permutations $S_{4n}$ acting on the equivalence classes of labeled graphs (comprising the encoded ring states), where two graphs are in the same equivalence class if they are graph isomorphisms of each other. For the encoded hexagonal ring state in Fig.~\ref{fig:24qubitgraph}, there are two graph isomorphisms: ($i$) the dihedral symmetry of the hexagon and ($ii$) the permutation symmetry of swapping the inner core and leaf qubits for the corresponding outer core and leaf qubits. Then, by the orbit-stabilizer theorem, the size of the orbit of the trivial emission ordering, and thus the total number of emission orderings, is the cardinality of $S_{4n}$ divided by the cardinality of the subgroup that leaves the graph invariant, which yields Eq.~\ref{eq:totalshorsize}. Note that this is still too large a search space to handle. For example, with the encoded hexagonal ring, this yields a search space size of $8.079\times10^{20}$. Indeed, this makes it computationally difficult to find the optimal generating circuits with brute force alone.

\begin{figure}[!htbp]
    \centering
    \includegraphics[width=0.5\textwidth]{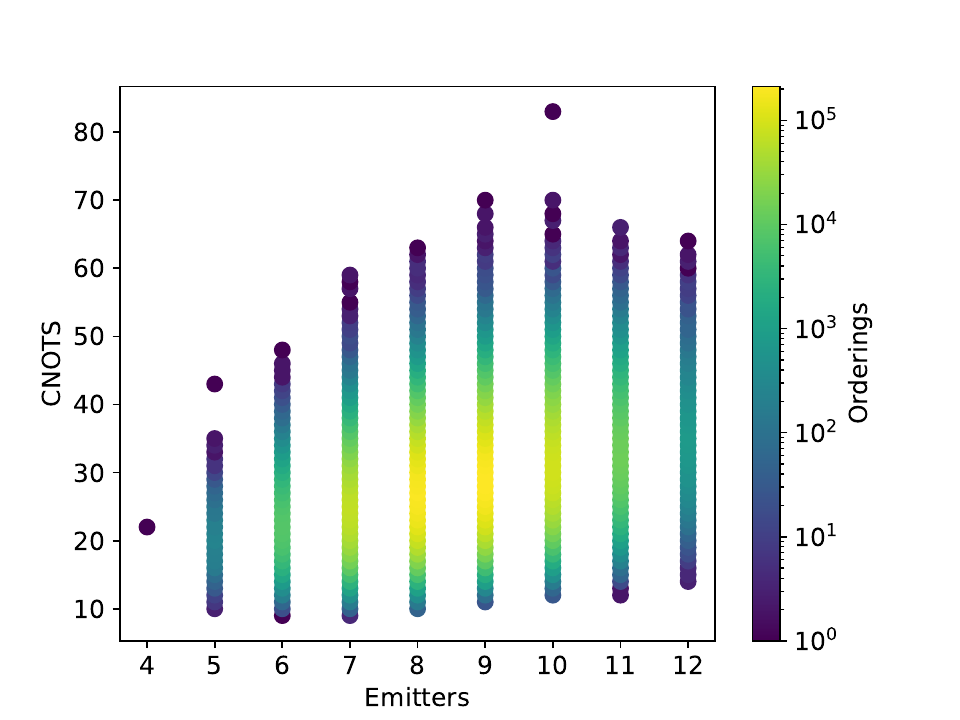}
    \caption{Emitter count vs CNOT count scatter plot of the (2,2) Shor-encoded 6-qubit ring for a randomized search. The color bar indicates the number of distinct emission orderings for each (Emitters, CNOTs) pair.}
    \label{fig:24QubitScatter}
\end{figure}

In light of this, one might instead consider performing a randomized search to find near-optimal generating circuits for encoded ring states. To test this idea, we randomly sampled a set of $N=4\times 10^{6}$ orderings for the encoded 6-ring state, in each case calculating the minimum number of emitters and the corresponding emitter-emitter CNOT count; the results are shown in  Fig.~\ref{fig:24QubitScatter}.
From these results, we find orderings that require as few as 4 emitters and 22 CNOTs. However, if the priority is to minimize the number of CNOTs instead, then one would either need 5 emitters and 10 CNOTs or 6 emitters and 9 CNOTs.

\begin{figure}[!htbp]
    \centering
    \includegraphics[width=0.3\textwidth]{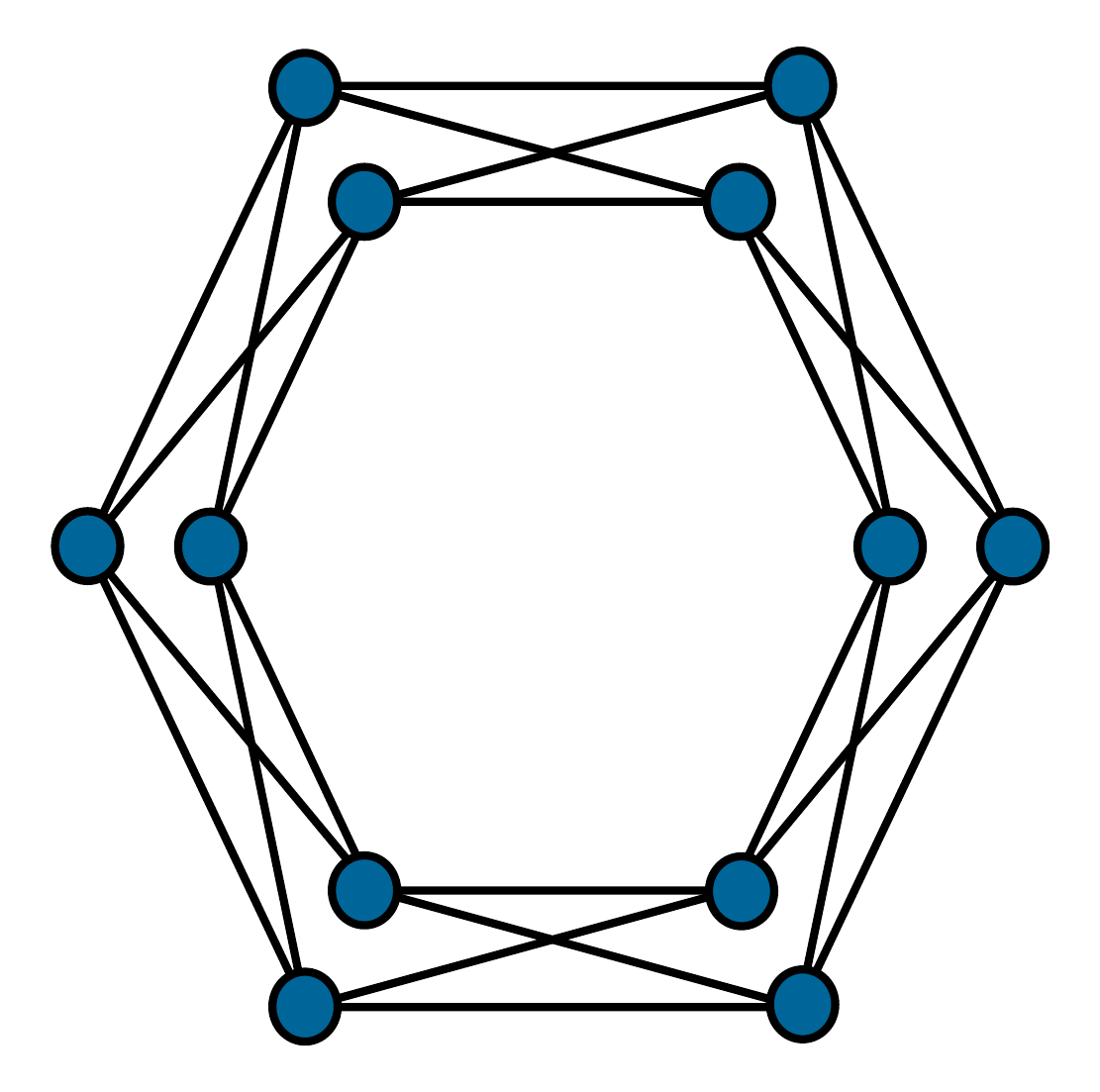}
    \caption{The 12-qubit simplified graph state of the original (2,2) Shor-encoded 6-ring resource state, where the leaf qubits have been removed.}
    \label{fig:12qubitgraph}
\end{figure}

These results can be improved further by exploiting graph truncation and symmetries, as we now show. We first note that the (2,2) Shor-encoding scheme contains $2n$ core qubits and $2n$ leaf qubits, where each core qubit is connected to one and only one leaf qubit. We propose simplifying the search through the following heuristic procedure: $(i)$ First, remove the leaf qubits. As a particular example, consider the 12-qubit graph in Fig.~\ref{fig:12qubitgraph} obtained from the original 24-qubit encoded graph in Fig.~\ref{fig:24qubitgraph} by removing its 12 leaf qubits. $(ii)$ Find the optimal emission orderings for the remaining, reduced graph of core qubits via a brute-force search (or a random search if $n$ is too large). $(iii)$ Add the leaf qubits back. In particular, we will restore the leaf qubits using the simple observation that once a core photon is emitted, we can emit its associated leaf photon immediately after from the same emitter. This means that, starting from an optimal ordering for the $2n$ core photons with labels $i=1,...,2n$ in the reduced graph, the core photons of the original $4n$-photon state will be labeled $i=1,3,...,4n-1$ (preserving the order), while the leaf qubits will be labeled $i=2,4,...,4n$ such that the leaf qubit with label $i$ is attached to the core photon labeled by $i-1$. $(iv)$ We then run the algorithm of Ref.~\cite{BikunsAlgorithm} on the $4n$-photon state with these emission orderings to obtain the associated, near-optimal circuits.

The (2,2) Shor-encoded version of an $n$-qubit ring will have a core of two interlocked $ n$-qubit rings with a total of $2n$ qubits. We can then compute the total number of distinct emission orderings for the $2n$-qubit interlocked ring state to find the size of the search space: 
\begin{eqnarray}
N = \dfrac{(2n)!}{n\cdot 2^{n+1}} \; .
    \label{eq:interlockedringsize}
\end{eqnarray}
This search space grows significantly slower than the search space computed in Eq.~\ref{eq:totalshorsize}. Whereas plugging $n=6$ into the formula of Eq.~\ref{eq:totalshorsize} gave a search space on the order of $10^{21}$, plugging in $n=6$ into the formula in Eq.~\ref{eq:interlockedringsize} yields $\frac{11!}{64}=623700$, which is a small enough search space to run a brute-force search. For larger resource states, the search space may be too large to run a brute-force search (such as $n=12$), but a randomized search will cover a larger portion of the search space at a significantly reduced computational cost.

\begin{figure}[!htbp]
    \centering
    \includegraphics[width=0.5\textwidth]{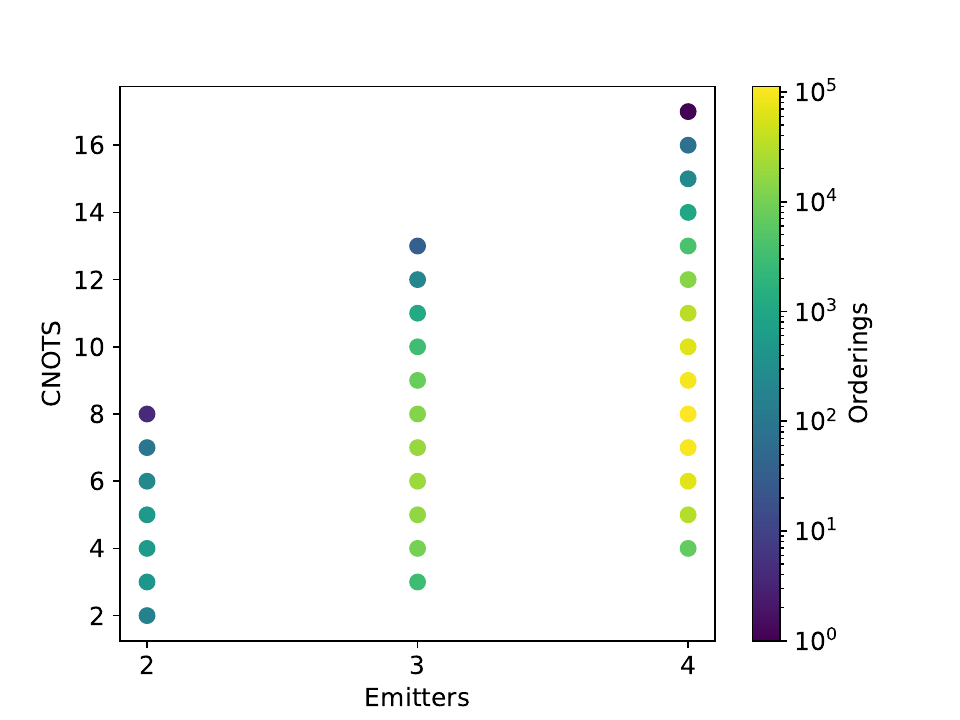}
    \caption{Emitter count vs CNOT count scatter plot of the encoded 6-qubit ring without its leaf qubits (see Fig.~\ref{fig:12qubitgraph}). The color bar indicates the number of distinct emission orderings for each (Emitters, CNOTs) pair.}
    \label{fig:12qubitscatter}
\end{figure}

Focusing on our example case of the encoded hexagonal ring, we obtain the scatter plot in Fig.~\ref{fig:12qubitscatter} by looking at every emission ordering for the 12-qubit interlocked ring in Fig.~\ref{fig:12qubitgraph}. We find that the optimal emission orderings for the interlocked rings have 2 emitters and 2 CNOTs between emitters. Furthermore, we find that there are 159 such emission orderings.

Next, we adopt the 159 optimized orderings from the interlocked 12-qubit ring and use them to generate orderings of the larger target 24-qubit state. We do this by applying the heuristic that we emit the leaf qubit either immediately before or after the emission of its associated core qubit (we check both cases and choose the one that leads to the smaller number of CNOTs). Effectively, this means that for each of the 159 optimal orderings we found for the truncated graph in Fig.~\ref{fig:12qubitgraph}, we produce a near-optimal ordering for the full graph by replacing each one-qubit emission by two, the core qubit and its leaf.

\begin{figure}[!htbp]
    \centering
    \includegraphics[width=0.5\textwidth]{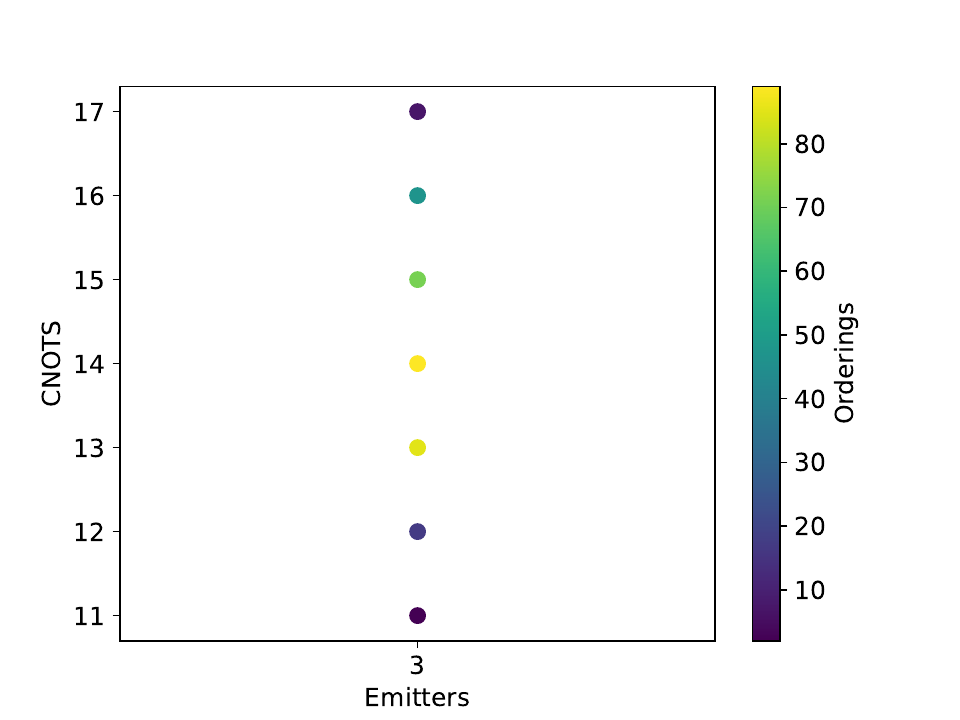}
    \caption{Emitter count vs CNOT count scatter plot of the (2,2) Shor-encoded 6-qubit ring after the reduced search. The color bar indicates the number of distinct emission orderings for each (Emitters, CNOTs) pair.}
    \label{fig:24QubitScatterOptimized}
\end{figure}

\begin{figure}
    \centering
    \includegraphics[width=0.5\textwidth]{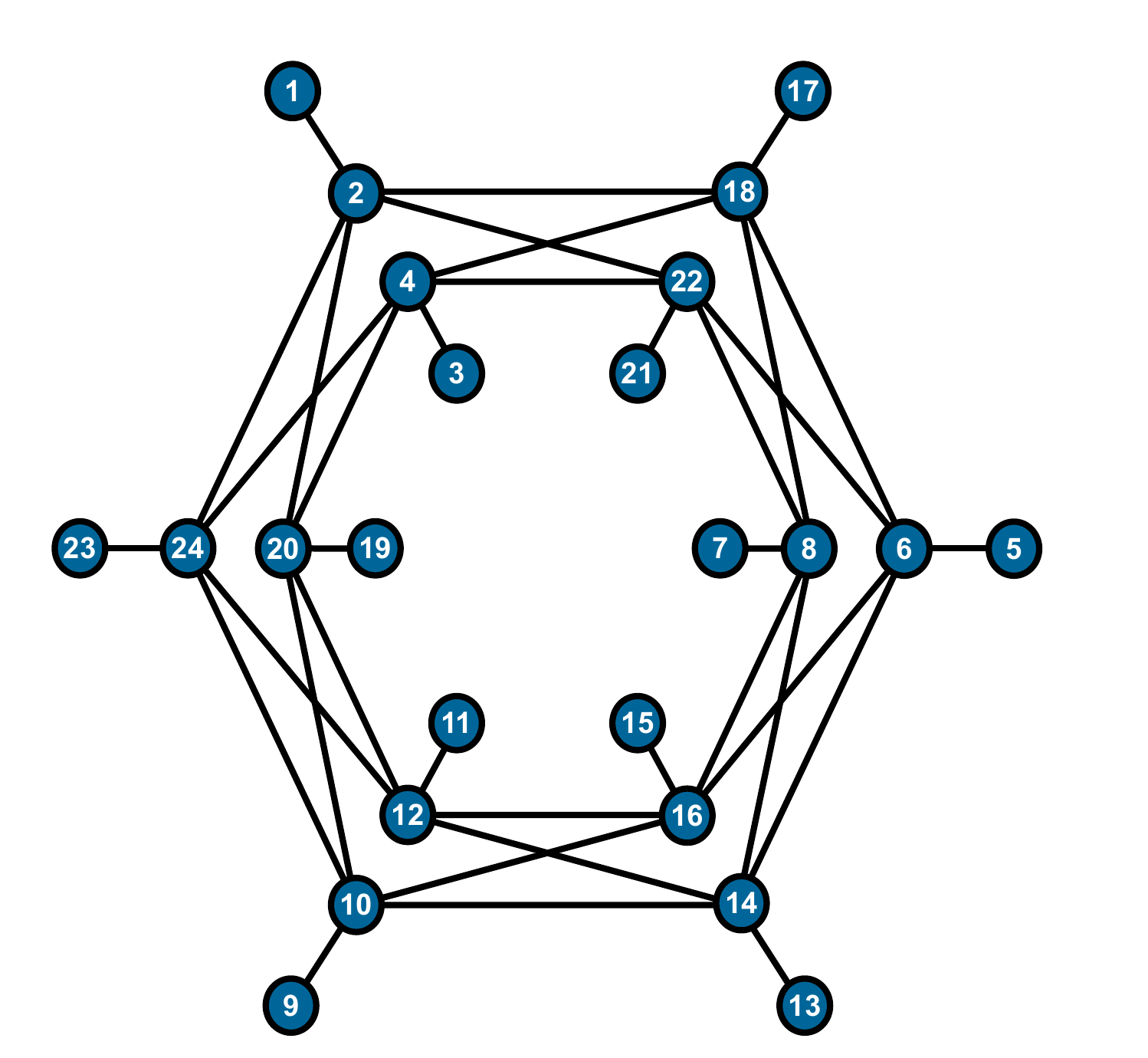}
    \caption{(2,2) Shor-encoded 6-ring resource state with optimized emission ordering.}
    \label{fig:24QubitLabeled}
\end{figure}

Using this heuristic, we obtain the scatter plot in Fig.~\ref{fig:24QubitScatterOptimized} for the emitter vs CNOT count. Note that with a significantly lower computational cost (compared to the randomized search approach in Fig.~\ref{fig:24QubitScatter}), we obtain a much better overhead of 3 emitters and 11 CNOTs. There are two such emission orderings, one of which is given in Fig.~\ref{fig:24QubitLabeled}. While this ordering is not guaranteed to be optimal, it is likely near-optimal given the low number of resources it requires. Indeed, it performs comparably in terms of the CNOT count and does better in terms of the emitter count relative to the two best orderings from the randomized search that yielded Fig.~\ref{fig:24QubitScatter}. Moreover, this near-optimal ordering is also not intuitive and unlikely to have been guessed upfront (e.g., photon 5 is not connected to 4, etc). For larger states, such as $n=12$, this approach should still be much more efficient at identifying near-optimal circuits compared to alternative methods. We also provide the generating circuit for this near-optimal ordering as follows:
\begin{widetext}
\begin{align}
U_{\text{circuit}} &= 
M_{25 \rightarrow 24}\, H_{24}\, H_{25}\, E_{25,24}\,
H_{23}\, H_{25}\, E_{25,23}\,
M_{26 \rightarrow 22}\, H_{22}\, H_{26}\, E_{26,22}\,
H_{21}\, H_{26}\, E_{26,21}\,
M_{27 \rightarrow 20}\, H_{20}\, H_{27}\, E_{27,20}
\nonumber \\
&\quad
H_{19}\, H_{27}\, E_{27,19}\,
H_{25}\, H_{27}\,
\mathrm{CNOT}_{27,25}\,
M_{25 \rightarrow 18}\,
H_{18}\, H_{25}\, E_{25,18}\,
H_{17}\, H_{25}\, E_{25,17}\,
H_{25}\, H_{26}\,
\mathrm{CNOT}_{26,25}\,
M_{25 \rightarrow 16}\,
\nonumber \\
&\quad
H_{16}\, H_{25}\, E_{25,16}\,
H_{15}\, H_{25}\, E_{25,15}\,
H_{15}\, H_{25}\,
H_{25}\,
\mathrm{CNOT}_{26,25}\,
\mathrm{CNOT}_{27,25}
M_{25 \rightarrow 14}\,
H_{14}\, H_{25}\,
E_{25,14}\,
\mathrm{CNOT}_{26,25}\,
\nonumber \\
&\quad
\mathrm{CNOT}_{27,25}\,
E_{25,13}\,
H_{25}\,
M_{25 \rightarrow 12}\,
H_{12}\, H_{25}\,
E_{25,12}\,
H_{11}\, H_{25}\,
E_{25,11}\,
H_{25}\, H_{27}\,
\mathrm{CNOT}_{27,25}\,
M_{25 \rightarrow 10}\,
H_{10}\, H_{25}\,
E_{25,10}\,
\nonumber \\
&\quad
H_{9}\, H_{25}\,
E_{25,9}\,
H_{25}\,
\mathrm{CNOT}_{27,25}\,
M_{25 \rightarrow 8}
H_{8}\, H_{25}\,
E_{25,8}\,
H_{26}\,
\mathrm{CNOT}_{26,25}\,
E_{25,7}\,
H_{26}\, H_{27}\,
\mathrm{CNOT}_{27,26}\,
E_{26,6}\,
H_{26}\,
E_{26,5}\,
\nonumber \\
&\quad
H_{25}\,
M_{25 \rightarrow 4}
H_{4}\, H_{25}\,
E_{25,4}\,
H_{27}\,
\mathrm{CNOT}_{27,25}\,
E_{25,3}\,
H_{27}\,
E_{27,2}\,
H_{1}\,
E_{27,1}\,
H_{25}\, H_{26}\, H_{27}
\; ,
\end{align}
\end{widetext}
where the photons are indexed $i=1, 2, \cdots, 24$ (corresponding to the order in which they are emitted) and the three emitters are denoted by the indices $i=25, 26, 27$, the gate $H_{i}$ is a Hadamard on the $i$-th qubit, $\mathrm{CNOT}_{i,j}$ denotes a CNOT gate between emitter $i$ (control qubit) and emitter $j$ (target qubit), $E_{i,j}$ denotes the emission of photon $j$ from emitter $i$ (equivalent to $\mathrm{CNOT}_{i,j}$ between them, where the photon is initialized in $\vert 0\rangle$), and $M_{i \rightarrow j}$ denotes the measurement of emitter $i$ in the Pauli $Z$ basis and feedforward onto photon $j$ an identity gate if the outcome is 0 and a Pauli $X$ gate if the outcome is 1. The initial input state of the circuit is assumed to be $\otimes_{i=1}^{27}\vert 0 \rangle$.

\section{Upper Bound on the Number of Emitter-Emitter CNOTs} \label{sec:bounds}
To complement the computational approach presented in the previous section, in this section, we present an upper bound on the total number of emitter-emitter CNOTs required in the generating circuits found by the algorithm of Ref.~\cite{BikunsAlgorithm}. We first note that the algorithm constructs the circuit in reverse, starting from the final state $\vert G \rangle \vert 0\rangle^{\otimes n_{e}}$ and ending with the fully disentangled state $\vert 0\rangle^{\otimes n_{p}} \vert 0\rangle^{\otimes n_{e}}$. We also refer to the algorithm steps by their time-reversed names to be consistent with Ref.~\cite{BikunsAlgorithm}. Hence, a single CNOT gate needed to entangle two emitters must belong to one of the three main steps described in the algorithm: (i) prior to each photon absorption (i.e., time-reversed photon emission), (ii) prior to each time-reversed emitter measurement, and (iii) at the end of the time-reversed circuit, where all photons are absorbed and only entangled emitters remain, e.g. $\vert 0 \rangle^{\otimes n_{p}}\vert S \rangle$ for some stabilizer state $S$ of the $n_{e}$ emitters. We denote the total number of CNOTs associated with photon absorptions by $\mathcal{A}$, the number associated with measurements by $\mathcal{M}$, and the number at the end of the circuit by $\mathcal{E}$. In Appendix~\ref{apx:main}, we obtain bounds for each of these contributions to the total CNOT account. The number of photon-absorption CNOTs obeys
\begin{eqnarray}
    \mathcal{A} \leq \left \lfloor \frac{n_{p}n_{e}}{2} \right \rfloor \; .
\end{eqnarray}
For the unencoded and encoded 6-ring, the r.h.s. yields 6 and 36 CNOTs, respectively. Furthermore, we show that the number of CNOT gates required to perform $n_{\text{TRM}}$ time-reversed measurements (where computing the value of $n_{\text{TRM}}$ requires running the algorithm) obeys 
\begin{equation}
    \mathcal{M} \leq  n_{\text{TRM}}\left \lfloor \frac{n_{e}}{2} \right \rfloor \leq n_{p}\left \lfloor \frac{n_{e}}{2} \right \rfloor \; ,
\end{equation}
which yields for the unencoded and encoded rings 6 and 24, respectively. Finally, the number of end-circuit CNOTs required to disentangle the emitters after all photons are absorbed obeys
\begin{equation}
    \mathcal{E} \leq \frac{n_{e}(n_{e}+1)}{2}- \left \lfloor \frac{(n_{e}+1)^{2}}{4} \right \rfloor \; .
\end{equation}
This yields 1 and 2 for the unencoded and encoded rings, respectively. Indeed, the sum of $\mathcal{A}$, $\mathcal{M}$, and $\mathcal{E}$ comprises an upper bound for the total number of emitter-emitter CNOT gates that we found in the previous section. This upper bound presumes that the weights of stabilizer generators are minimized at each step as in Ref.~\cite{takou2025optimization} and yields an upper bound of 13 and 62 for the total number of emitter-emitter CNOTs in the generating circuits of the unencoded and encoded 6-ring states, respectively. Although for these specific examples, these upper bounds are loose, they can still provide a useful point of reference for more complicated target graph states.

\section{Discussion and Conclusion} \label{sec:conclusion}
In this work, we identified near-optimal generating circuits for photonic resource states for fault-tolerant FBQC. By exploiting three characteristic features of the target states—namely their small size, symmetry, and graph structure, including the presence of leaf qubits—we reduced the search space sufficiently to make a brute-force strategy practical. This allowed us to identify near-optimal, and in some cases plausibly optimal, generating circuits as measured by emitter number and emitter-emitter CNOT count. Applied to the 24-photon Shor-encoded resource state of Ref.~\cite{FBQC}, this approach yields a construction requiring only 3 emitters and 11 emitter-emitter CNOT gates. This remarkably compact circuit suggests that the resource requirements of emitter-based photonic state generation may be substantially lower than previously appreciated. The significance of this reduction is not limited to only the scale of the requisite hardware. A smaller emitter register also shortens the time over which emitters must remain entangled with the photons during state generation, thereby reducing exposure to errors, especially those originating from emitter spin decoherence. Likewise, using fewer emitters may ease requirements on emitter homogeneity, including spectral matching and frequency stability, which become increasingly demanding as the number of emitters grows. 

Our results connect naturally to recent proposals based on hybrid deterministic-probabilistic architectures. In addition to fully deterministic generation schemes, there exist approaches that combine deterministic state generation at the emitter level with probabilistic photonic fusion operations~\cite{hilaire2023near, wein2025minimizing, chan2025tailoring,meng2025temporal}. For example, Ref.~\cite{wein2025minimizing} proposed near-deterministic generation of the Shor-encoded (2,2) 6-ring resource state by fusing deterministically generated caterpillar states \cite{hilaire2023near}. Such an approach has clear practical appeal: caterpillar states can be generated from a single emitter and then used as modular building blocks for boosted fusion. In hybrid schemes that involve fusing states that are more complicated than caterpillars and require more than one emitter to generate, the compact constructions identified here suggest that the underlying deterministic component of such hybrid schemes may itself admit further reduction.

It is also important to emphasize that our results are not tied to any particular hardware implementation; the required emitter-emitter entangling operations could be realized through direct or cavity-mediated interactions, photon-mediated repeat-until-success (RUS) gates~\cite{RUS2004}, or hybrid combinations thereof, depending on the platform. This point is particularly relevant in light of Ref.~\cite{wein2025minimizing}, which showed that a 12-emitter architecture based on 14 photon-mediated RUS entangling gates can already substantially outperform probabilistic linear-optical approaches once loss and device nonidealities are taken into account. Our results suggest that there may still be considerable room to reduce the overhead of such emitter-based schemes at the circuit level. Specifically, our framework makes it possible to optimize over constructions involving between 3 and 12 emitters, opening the door to systematic trade-off studies between emitter number, entangling-gate count, entanglement-storage time, and architectural constraints. In this sense, the present work should be viewed not only as an explicit compact construction for one important FBQC resource state, but also as a general method for exploring the design space of emitter-based photonic resource-state generation. How the reductions enabled by our work translate into architecture-level gains will depend on the physical implementation, since deterministic interactions and RUS gates introduce different trade-offs in fidelity, timing, and optical loss. Even so, the fact that the same target state can be realized with so few emitters and entangling gates strengthens the case that emitter-based generation remains highly competitive, and perhaps even more promising than previously recognized.

Beyond fault-tolerant computing, these ideas are also relevant to other photonic quantum technologies. For example, Ref.~\cite{pettersson2025long} proposed encoded ring states for quantum communication using concatenated graph-code constructions~\cite{bell2023optimizing}, where the number of matter qubits scales with the level of concatenation. While such encodings provide useful loss tolerance for logical Pauli measurements and fusion operations, they are unlikely to be optimal with respect to emitter number. This suggests another promising application of the methods developed here: searching for more emitter-efficient generating circuits for encoded graph states used in quantum networking and quantum sensing, where resource states also tend to possess a high degree of symmetry.

Our results strengthen the case for emitter-based resource-state generation as an alternative to architectures based solely on single-photon sources and linear optics and show that resource-state generation admits a much richer optimization landscape than has so far been exploited. As emitter platforms improve in two-qubit gate fidelity and optical interface efficiency, the advantages of compact constructions are likely to become increasingly pronounced. A full architecture-level analysis including loss, finite-fidelity operations, scheduling constraints, and the choice between deterministic, probabilistic, and hybrid entangling mechanisms will be needed to quantify the relative merits and ultimately the optimal approach  in a given platform. We leave such analyses to future work.

\section{Acknowledgements}
The authors thank Tarek Razzaz for useful discussions. SEE acknowledges support from the US Department of Energy, Office of Basic Energy Sciences, through the
Quantum Photonic Integrated Design Center (QuPIDC) EFRC award DE-SC0025620. EB acknowledges support from the National Science Foundation, grant no. 2137953. 

\section{Code Availability}
The Python code for implementing the circuit-solving algorithm \cite{BikunsAlgorithm} can be found at \url{https://github.com/nrmanohar/photonic_circuit_solver}

\bibliography{references}

\appendix

\section{Theoretical Analysis of the Algorithm} \label{apx:main}
\subsection{RREF Gauge of a Generator Set}
Consider the list of $n$ generators $\mathcal{G}$, and denote by $l(g)$ the qubit index corresponding to the leftmost non-trivial Pauli operator of the generator $g \in \mathcal{G}$. Then, the RREF gauge \cite{RREF} (after transforming $\mathcal{G}$ with appropriate operations that preserve the stabilizer set) is written as a set of $q\leq n$ generators with increasing leftmost Pauli index $l_{1}<l_{2}<\cdots<l_{q}$, where the remaining $p=n-q$ generators have leftmost Pauli indices given by an ordered sublist $l_{\mu_{1}}<l_{\mu_{2}}<\cdots<l_{\mu_{p}}$ of the ordered list $\{l_{i}\}_{i=1}^{q}$, hence $p\leq q$. Therefore, the equivalent set of generators in the RREF gauge is given by 
\begin{eqnarray}
    \mathcal{G}_{\text{RREF}}=\{g_{l_{1}}, g_{l_{2}}, \cdots, g_{l_{q}}\}\cup \{g_{l_{\mu_{1}}}^{\prime}, g_{l_{\mu_{2}}}^{\prime}, \cdots, g_{l_{\mu_{p}}}^{\prime}\} \; .
\end{eqnarray}
By construction, the generating set $\mathcal{G}_{\text{RREF}}$ has no more than two generators with the same left-most Pauli index. Part of the definition of the RREF gauge is that for any pair of generators with the same left-most index $l_{\mu_{i}}$ for $i=1, 2, \cdots, p$, multiplying them together does not change the left-most Pauli index, i.e., $l\left(g_{l_{\mu_{i}}}g_{l_{\mu_{i}}}^{\prime}\right)=l_{\mu_{i}}$. This means that the two left-most Pauli operators cannot be identical, and hence must anti-commute.

\subsection{RREF Gauge in the Li \textit{et al.} Algorithm \cite{BikunsAlgorithm}}
The algorithm proposed in Ref.~\cite{BikunsAlgorithm} is used to find the generating circuit of an arbitrary photonic graph state of $n_{p}$ photons and a minimal number of $n_{e}$ emitters. Given a target photonic graph state and emission ordering (specified by labeling the vertices of the graph by indices $j=1,2,\ldots,n_{p}$, where $j=1$ corresponds to the first emitted photon), the algorithm first determines $n_{e}$. It then works backward in time-reversed order, starting from the target graph state and $n_e$ decoupled emitters, in each step either absorbing a photon into an emitter or performing a time-reversed measurement on an emitter in preparation for the next photon absorption. Namely, the algorithm runs so that at step $j=n_{p}$ (the first step), the $n_{p}$-th photon is absorbed (disentangled from the graph). Then, at the next step $j=n_{p}-1$, the $(n_{p}-1)$-st photon is absorbed, and so on, until at step $j=1$ (last step), the 1st photon is absorbed. Therefore, at any step $j$ of the algorithm, we would have the photons $n_{p}, n_{p}-1, \cdots, j+1$ disentangled from the graph. By noting that, in the algorithm \cite{BikunsAlgorithm}, the state of the photons and the emitters is a stabilizer state at each step, where the generators have the form: $g=\sigma_{1}\otimes \sigma_{2} \otimes \cdots \sigma_{n_{p}}\otimes \tau_{1}\otimes \tau_{2} \otimes \cdots \otimes \tau_{n_{e}}$, where the Pauli operators $\{\sigma_{i}\}_{i=1}^{n_{p}}$ of the emission-ordered photons appear first, followed by the emitter Pauli operators $\{\tau_{i}\}_{i=1}^{n_{e}}$ (in no particular order). Therefore, we would have a stabilizer state with the generators in the RREF gauge given by 
\begin{align}
    \mathcal{G}_{\text{RREF}}=\mathcal{G}^{(j)}_{\text{RREF}} \cup \{Z_{j+1}, \cdots, Z_{n_{p}}\} \; ,
\end{align}
with 
\begin{equation}
    \mathcal{G}_{\text{RREF}}^{(j)}=\mathcal{G}^{\leq j}_{\text{RREF}} \cup \mathcal{G}^{>n_{p}}_{\text{RREF}} \; ,
\end{equation}
where $\mathcal{G}^{\leq j}_{\text{RREF}} $ and $\mathcal{G}^{>n_{p}}_{\text{RREF}}$ denote the subset of generators with left-most Pauli indices smaller than $j$ or larger than $n_{p}$, respectively. The stabilizers $Z_{j+1}, \ldots, Z_{n_p}$ correspond to the photons that have already been absorbed at this stage of the algorithm, as we mathematically represent an absorbed photon using the state $\ket{0}$. We denote by $\left\vert \mathcal{G}_{\text{RREF}}^{\leq j} \right\vert \equiv a$ the number of generators with $l(g)\leq j$, and hence we have $\left\vert \mathcal{G}^{>n_{p}}\right\vert=n_{e}+j-a$ being the number of generators that have support only on the emitters, consistent with $\left\vert \mathcal{G}_{\text{RREF}}^{(j)} \right\vert=n-(n_{p}-j)=n_{e}+j$. Therefore, the RREF gauge at step $j$ looks like
\begin{equation}
\left[
    \begin{array}{c|c||c}
        \begin{array}{c}
             \text{RREF}_{a \times j}
        \end{array} & I_{a \times (n_{p}-j) }& \cdots \\
        \hline 
        I_{(n_{p}-j) \times j} & 
        \begin{array}{ccc}
             Z_{j+1} &  & \\
             & \ddots & \\
             &  & Z_{n_{p}}
        \end{array} & I_{(n_{p}-j) \times n_{e}} \\
        \hline 
        I_{(n_{e}+j-a) \times j} & I_{(n_{e}+j-a) \times (n_{p}-j) } & 
        \begin{array}{c}
             \text{RREF}_{(n_{e}+j-a)\times n_{e}} 
        \end{array}
    \end{array}
\right]
\end{equation}
We note that the number of generators in $\mathcal{G}_{\text{RREF}}^{>n_{p}}$ must not overdetermine the state of the emitters, i.e., the number of generators in $\text{RREF}_{(n_{e}+j-a) \times n_{e}}$ should not exceed $n_{e}$. This implies that $\left\vert \mathcal{G}_{\text{RREF}}^{>n_{p}} \right\vert \leq n_{e}$, and hence $a \ge j$. However, because the total number of generators with no support on the absorbed photons at step $j$ is given by $n_{e}+j$, it follows that $a \leq \min \{2j, j+n_{e}\}$, where $a=2j$ is the case where all generators with indices smaller or equal to $j$ have repetitions.

\subsection{Bipartite Entanglement of a Stabilizer State in the RREF Gauge}
Consider the general case of $n$ independent generators $\mathcal{G}\equiv \mathcal{G}_{l<r} \cup \mathcal{G}_{l\ge r}$, where $\vert \mathcal{G}_{l \ge r}\vert = m$ of them have $l(g) \ge r$ and $\vert \mathcal{G}_{l<r}\vert = n-m$ of them have $l(g)<r$. We organize the generators as follows:
\begin{align}
    g_{i} &=II\cdots I\sigma^{(i)}_{r}\sigma^{(i)}_{r+1}\cdots \sigma^{(i)}_{n} \hspace{0.2cm} \text{for} \hspace{0.2cm} i=1, \cdots, m \; , \label{eqn:gen_limted} \\
    g_{i} &= \sigma_{1}^{(i)}\cdots \sigma^{(i)}_{r}\sigma^{(i)}_{r+1}\cdots \sigma^{(i)}_{n} \hspace{0.2cm} \text{for} \hspace{0.2cm} i=m+1, \cdots, n \; . \label{eqn:gen_full}
\end{align}
Our goal is to find out under what conditions the corresponding stabilizer state can become separated between the aforementioned $m$ and $n-m$ qubits.

We start by writing the stabilizer state $\vert \chi \rangle$ corresponding to this generating set in the Schmidt decomposition (where the bipartition is performed with respect to the two qubit subsets $F\equiv \{1, \cdots, r-1\}$ and $H \equiv \{r, \cdots, n\}$) as
\begin{align}
    \vert \chi \rangle = \sum_{\alpha=1}^{s}\sqrt{\lambda_{\alpha}}\vert f_{\alpha}\rangle_{F} \otimes \vert h_{\alpha}\rangle_{H} \; ,
\end{align}
where $s$ is the Schmidt number, and $\{\lambda_{\alpha}\}_{\alpha=1}^{s}$ comprise a normalized probability distribution. By multiplying the stabilizer state $\vert \chi \rangle$ with an arbitrary stabilizer of the form $\sigma_{1}^{(k)}\cdots \sigma_{n}^{(k)}$ and taking the overlap with the Schmidt component $\langle f_{\beta}\vert \otimes \langle h_{\beta} \vert$ with an arbitrary $\beta \in \{1,\ldots,s\}$, we arrive at 
\begin{align}
    \sum_{\alpha=1}^{s}\sqrt{\frac{\lambda_{\alpha}}{\lambda_{\beta}}}\langle f_{\beta} \vert \sigma_{1}^{(k)}\cdots \sigma_{r-1}^{(k)}\vert f_{\alpha} \rangle \langle h_{\beta} \vert \sigma_{r}^{(k)}\cdots \sigma_{n}^{(k)}\vert h_{\alpha}\rangle = 1 \; . \label{eqn:Schmidt}
\end{align}
For all $k \leq m$ (i.e., $\forall g \in \mathcal{G}_{l\ge r}$), we have $\sigma_{1}^{(k)}\cdots \sigma_{r-1}^{(k)}=I\cdots I$, which yields for Eq.~\eqref{eqn:Schmidt}
\begin{equation}
    \langle h_{\alpha} \vert \sigma_{r}^{(k)}\cdots \sigma_{n}^{(k)}\vert h_{\alpha}\rangle = 1 \; ,
\end{equation}
for all $\alpha=1, \cdots, s$. This condition reads equivalently
\begin{equation}
    \sigma_{r}^{(k)}\cdots \sigma_{n}^{(k)}\vert h_{\alpha}\rangle = \vert h_{\alpha} \rangle \; ,
\end{equation}
for all $\alpha=1, \cdots, s$. In other words, the Schmidt states $\vert h_{\alpha} \rangle$ belong to the codespace of the $m$ generators given by Eq.~\ref{eqn:gen_limted}. Since the number of generators $g_{1}, g_{2}, \cdots, g_{m}$ must be smaller than or equal to the number of qubits comprising the states $\vert h_{\alpha}\rangle$, we have $m \leq n-r+1$, where the equality means that $\vert h_{\alpha}\rangle \equiv \vert h\rangle$ is fully determined to be the unique stabilizer state of the generating set in Eq.~\eqref{eqn:gen_limted}, which yields $\vert \chi \rangle = \vert f\rangle_{F}\otimes \vert h\rangle_{H}$, i.e., $s=1$. In the opposite case of $m > n-r+1$, it must mean that some of the generators in Eq.~\eqref{eqn:gen_limted} are not independent, which contradicts our assumption. In general, the size of the codespace defined by the subset of generators in Eq.~\eqref{eqn:gen_limted} is given by $2^{(n-r+1)-m}$, which is the largest that the Schmidt number $s$ can be for the basis $\vert h_{\alpha}\rangle$ with $\alpha=1, \cdots, s$ to remain mutually orthogonal.

Now we consider the rest of the generators $k = m+1, \cdots, n$ (i.e., $\forall g \in \mathcal{G}_{l< r}$). We first start by noting that in the term $ \langle h_{\beta} \vert \sigma_{r}^{(k)}\cdots \sigma_{n}^{(k)}\vert h_{\alpha}\rangle$ in Eq.~\eqref{eqn:Schmidt}, the $\sigma_{r}^{(k)}\cdots \sigma_{n}^{(k)} \equiv \hat{g}_{k}$ part of the generator $g_{k}=\sigma_{1}^{(k)}\cdots \sigma_{n}^{(k)}\equiv (\overline{g}_{k}\vert \hat{g}_{k})$ commutes with the previous $m$ generators $g_{t}=\sigma_{r}^{(t)}\cdots \sigma_{n}^{(t)}$ for $t=1, \cdots, m$. Therefore, $\hat{g}_{k}=\sigma_{r}^{(k)}\cdots \sigma_{n}^{(k)}$ belongs to the centralizer $\mathcal{C}(\mathcal{S}_{l \ge r})$ of the stabilizer set $\mathcal{S}_{l \ge r}=\langle \mathcal{G}_{l\ge r}\rangle=\langle \{g_{t}\}_{t=1}^{m}\rangle $. One of two things is possible: if $\hat{g}_{k}$ belongs to the stabilizer set itself $\mathcal{S}_{l \ge r}$, then it could be written as the product of its generators $\{g_{t}\}_{t=1}^{m}$ (we can ignore this case w.l.o.g. because we can then define a new generator $g_{k}^{\prime}$ after multiplying $g_{k}=(\overline{g}_{k}\vert \hat{g}_{k})$ by the appropriate generators $\{g_{t}\}_{t=1}^{m}$ to remove the $\hat{g}_{k}$ part, which yields $g^{\prime}_{k}=(\overline{g}_{k} \vert I \cdots I)$, without impacting the leftmost Pauli index $l(g)$ of $g_{k}$). The only alternative is that $\hat{g}_{k} \in \mathcal{L}_{l \ge r} = \mathcal{C}(\mathcal{S}_{l \ge r})/\mathcal{S}_{l \ge r}$ belongs to the set of logical operators of the codespace generated by $\{g_{t}\}_{t=1}^{m}$. In this case, $ \sigma_{r}^{(k)}\cdots \sigma_{n}^{(k)}\vert h_{\alpha} \rangle$ would change the state $\vert h_{\alpha}\rangle$ to a different state in the same codespace of dimension $2^{(n-r+1)-m}$. Therefore, the term $ \langle h_{\beta} \vert \sigma_{r}^{(k)}\cdots \sigma_{n}^{(k)}\vert h_{\alpha}\rangle$ in Eq.~\eqref{eqn:Schmidt} is not zero if and only if $\vert h_{\alpha}\rangle$ and $\vert h_{\beta}\rangle$ are connected by the logical operation $\hat{g}_{k}$. The number of such logical operations in the codespace spanned by $\mathcal{S}_{l \ge r}$ is given by $2[(n-r+1)-m]$, corresponding to the single-qubit Pauli $X$ and $Z$ acting on $(n-r+1)-m$ logical qubits. 
The minimum weight of the logical operators $\hat{g}_{k}$ for $k=m+1, \cdots, n$ is then determined by the code distance $d$ of the stabilizer code $\mathcal{S}_{l \ge r}$. Then, the quantum Singleton bound provides an upper bound to the code distance $[(n-r+1)]-[(n-r+1)-m] \ge 2(d-1)$, which yields $d \leq \frac{m}{2}+1$. Therefore, the weight of the smallest generator from $\mathcal{G}_{l<r}$ (with a single Pauli operator at qubit index less than $r$) is lower bounded by $d+1 \leq \frac{m}{2}+2$.

\subsubsection{Photon Absorption}
We recall that, at the $j$-th step of photon absorption, there exists a stabilizer with $l(g)=j$ of the form $g_{a}=(Z_{j}\vert\hat{g}_{a})$ \cite{BikunsAlgorithm}, which is the generator with the largest $l(g)$ in $\mathcal{G}_{\text{RREF}}^{\leq j}$. Therefore, the number of CNOTs required to disentangle the emitters before photon absorption is given by $d-1\leq \frac{m}{2}$ (which is equal to one less than the weight of $\hat{g}_{a}$). This presumes that we are performing weight minimization of the given generator by multiplying it by the remaining generators to bring its weight to $d$ (this was proposed recently in Ref.~\cite{takou2025optimization}). For $r=n_{p}+1$ (the first emitter index) and $m=n_{e}+j-a$, we have for the upper bound $\frac{n_{e}+j-a}{2}$. Recalling that $j \leq a(j) \leq \min \{2j, j+n_{e}\}$, we have for the total number of emitter-emitter CNOTs during photon absorption (which we denote by $\mathcal{A}$) the following:  
\begin{align}
    \mathcal{A} & \coloneqq \sum_{j=1}^{n_{p}}[wt(g_{j})-1] \\ 
    &\leq \sum_{j=1}^{n_{p}}\frac{n_{e}+j-a}{2} \\
    &\leq \sum_{j=1}^{n_{p}} \frac{n_{e}}{2} = \frac{n_{p}n_{e}}{2} \; ,
\end{align}
which implies (because $\mathcal{A}$ is an integer) the upper bound
\begin{eqnarray}
    \mathcal{A} \leq \left \lfloor \frac{n_{p}n_{e}}{2} \right\rfloor  \; .
\end{eqnarray}

\subsubsection{Time-Reversed Measurement}
We note that the number of non-photonic generators $\mathcal{G}_{\text{RREF}}^{\ge n_{p}}$ is given by $n_{e}+j-a$. Therefore, by using $j \leq a \leq \min \{2j, j+n_{e}\}$, the number of non-photonic generators is in the range $[\max \{n_{e}-j, 0\}, n_{e}]$. This implies that the minimum weight of a non-photonic generator is bounded from above by $wt(g_{l_{\text{max}}}) \leq \left \lfloor \frac{n_{e}}{2} \right \rfloor + 1$, where $g_{l_{\text{max}}} \in \mathcal{G}_{\text{RREF}}^{\ge n_{p}}$ is the generator with the largest left-most Pauli index $l(g)=l_{\text{max}}$. This yields the total number of emitter-emitter CNOTs required to perform the time-reversed measurement (TRM) in the algorithm \cite{BikunsAlgorithm} as one less than the weight of the smallest non-photonic generator (based on our heuristic, we always choose the non-photonic generator with the lowest weight for the TRM, which may not be unique). Therefore, for a total of $n_{\text{TRM}}$ required TRMs (computed from the algorithm in Ref.~\cite{BikunsAlgorithm} for a given graph), we arrive at the upper bound for the total number of  CNOTs required to perform all TRMs (which we denote by $\mathcal{M}$) as:
\begin{equation}
    \mathcal{M} \leq n_{\text{TRM}}\left \lfloor \frac{n_{e}}{2} \right \rfloor \; .
\end{equation}
Since $n_{\text{TRM}}$ is bounded from above by the number of absorption steps $n_{p}$, it follows that the $n_{\text{TRM}}$-agnostic upper bound reads $\mathcal{M} \leq n_{p}\left \lfloor \frac{n_{e}}{2} \right \rfloor$.

\subsubsection{End-Circuit Measurements}
To fully account for the total number of CNOTs required by the algorithm in Ref.~\cite{BikunsAlgorithm}, we still need to account for the number of CNOTs consumed to disentangle the emitters at the end of the algorithm (i.e., after all photons are absorbed). To derive an upper bound on the number of these CNOT gates, we note that the Pauli weights of the non-photonic generators $\{g_{i}\}_{i=1}^{n_{e}}$ (i.e., $l(g_{i})>n_{p}$) are bounded from above in the RREF gauge by $wt(g_{i})\leq n_{e} - \lceil \frac{i}{2} \rceil + 1 $ (all photonic generators, given by $l(g) \leq n_{p}$, are of the form $Z_{j}$ for photons $j=1, \cdots, n_{p}$, since they are all disentangled). Therefore, the number of CNOTs required to disentangle one emitter from the rest within $g_{1}$ is given by $wt(g_{1})-1$. Once one emitter is disentangled, the weight of the rest of the emitter generators is reduced by one, by appropriate multiplication with $g_{1}$. By continuing to disentangle emitters one by one, moving from $g_{1}$ to $g_{2}$ to $g_{3}$ etc., it follows that the total number of CNOTs for an end-circuit measurement and resetting (which we denote by $\mathcal{E}$) is upper bounded by
\begin{align}
    \mathcal{E} &\coloneqq \sum_{i=1}^{n_{e}}(wt(g_{i})-i) \\ & \leq n_{e}^{2}-\sum_{i=1}^{n_{e}}\lceil \frac{i}{2} \rceil +n_{e} -\sum_{i=1}^{n_{e}}i \\ 
    &= n_{e}^{2} + n_{e} - \left \lfloor \frac{(n_{e}+1)^{2}}{4} \right \rfloor - \frac{n_{e}(n_{e}+1)}{2} \\ 
    &=\frac{n_{e}(n_{e}+1)}{2}- \left \lfloor \frac{(n_{e}+1)^{2}}{4} \right \rfloor \; .
\end{align}

\end{document}